\documentstyle[art12,bezier]{article}

\newcommand\sect[1]{\setcounter{equation} 0\section{#1}}

\renewcommand\theequation{\thesection\arabic{equation}}   

\newcommand{\be}{\begin{equation}}
\newcommand{\ee}{\end{equation}}
\newcommand{\ba}{\begin{eqnarray}}
\newcommand{\ea}{\end{eqnarray}}
\newcommand{\baa}{\begin{eqnarray*}}
\newcommand{\eaa}{\end{eqnarray*}}
\newcommand{\bb}{\begin{thebibliography}}
\newcommand{\eb}{\end{thebibliography}}
\newcommand{\ci}[1]{\cite{#1}}
\newcommand{\bi}[1]{\bibitem{#1}}
\newcommand{\lab}[1]{\label{#1}}             
\newcommand{\re}[1]{(\ref{#1})}
\newcommand{\Tr}{\mbox{tr\,}}

\def\Re{\mbox{Re}}

\def\IR{infrared}

\def\e{\mbox{e}}
\def\ie{\hbox{\it i.e.}{ }}
\def\CO{{\cal O}}
\def\CP{{\cal P}}
\def\CT{{\cal T}}

\def\CG{{\cal G}}

\newcommand\qqqquad{\qquad\qquad}

\newcommand\vev[1]{\langle{#1}\rangle}

\newcommand\fracs[2]{\mbox{\small $\frac{#1}{#2}$}}
\newcommand\cc[1]{\frac{\alpha_s}{{#1}\pi}} 
\newcommand\ccc[1]{\left(\cc{}\right)^{#1}}
\newcommand\lr[1]{\left({#1}\right)}

\def \as{\relax\ifmmode\alpha_s\else{$\alpha_s${ }}\fi}
\def \alpi {\frac \as \pi}

\def\MS{\overline{MS}}

\begin{document}
\renewcommand{\thefootnote}{\fnsymbol{footnote}}

\thispagestyle{empty}

\hfill\parbox{45mm}{
{\sc DFPD--93/TH/23}  \par
{\sc UPRF--93--366}   \par
{\sc UTF--93--291}    \par
{\sc hepph@xxx/9303314 } \par
March 1993}

\vspace*{15mm}

\begin{center}
{\LARGE Gauge Invariance and Anomalous Dimensions of a Light-Cone
        Wilson Loop in Light-Like Axial Gauge}

\vspace{22mm}

{\large A.~Bassetto}

\medskip

{\em Dipartimento di Fisica, Universit\`a di Padova and \par
INFN, Sezione di Padova, 35100 Padova, Italy}

\bigskip

{\large I.~A.~Korchemskaya}%
\footnote{On leave from the Moscow Energetic Institute, Moscow, Russia}
,\qquad
{\large G.~P.~Korchemsky}%
\footnote{On leave from the Laboratory of Theoretical Physics,
          JINR, Dubna, Russia}

\medskip

{\em Dipartimento di Fisica, Universit\`a di Parma and \par
INFN, Gruppo Collegato di Parma, 43100 Parma, Italy}

\bigskip

and

\bigskip

{\large G.~Nardelli}

\medskip

{\em Dipartimento di Fisica, Universit\`a di Trento and \par
INFN, Gruppo Collegato di Trento, 38050 Povo (Trento), Italy}

\bigskip

\medskip

\end{center}

\vspace*{20mm}

\renewcommand{\thefootnote}{\arabic{footnote}}
\setcounter{footnote} 0
\begin{abstract}
Complete two-loop calculation of a dimensionally
regularized Wilson loop with light-like segments
is performed in the light-like axial gauge with
the Mandelstam-Leibbrandt prescription for the gluon
propagator. We find an expression which {\it exactly}
coincides with the one previously obtained for the same Wilson
loop in covariant Feynman gauge.
The renormalization of Wilson loop is performed in the
$\MS-$scheme using a general procedure tailored
to the light-like axial gauge. We find that the
renormalized Wilson loop obeys a renormalization group equation
with the same anomalous dimensions as in covariant gauges.
Physical implications of our result for investigation of
infrared asymptotics of perturbative QCD are pointed out.
\end{abstract}

\newpage

\sect{Introduction}
There exists a unique possibility in non-Abelian gauge theories to
formulate the nontrivial dynamics of gauge fields in terms of the
gauge invariant collective variables \ci{Mig}
$$
W_C=\frac1N\vev{0|\ \Tr\CT\CP\exp\left(ig\oint_C dx^\mu
A_\mu^a(x)t_a\right)|0},
$$
called the Wilson loop expectation values. Here, $A_\mu^a(x)$ is the gauge
field, $\CT$ orders gauge field operators in time and $\CP$ orders
generators $t_a$ of the $SU(N)$ gauge group along a closed integration
path $C.$ With this definition, $W_C$ is a gauge invariant non local
functional of a gauge field depending on the integration path and
satisfying nontrivial loop equations \ci{Mig}.

However, a closer consideration shows \ci{Pol,WL} that $W_C$ is not
a well defined quantity for some paths $C$ lying in $4-$dimensional
Minkowski space-time. The problem is that for a closed path $C$ having
either cusps, or self-intersections \ci{Pol}, or segments lying on the
light-cone \ci{WL}, $W_C$, when calculated in perturbation theory, is
divergent even after renormalization of gauge field and coupling constant.

Perturbation theory expansion of $W_C$ looks like
\be
W_C=1+\sum_{n=2}^\infty (ig)^n\oint_C dx_1^{\mu_1}\cdots\oint_C dx_n^{\mu_n}
    \theta(x_1>\cdots>x_n)\frac1N\Tr G_{\mu_1\cdots\mu_n}(x_1,\ldots,x_n),
\lab{def}
\ee
where $G_{\mu_1\cdots\mu_n}(x_1,\ldots,x_n)$ is the $n-$point
Green function and the $\theta-$function orders points $x_1,\ldots,x_n$
along the integration path $C.$

Let us consider the simplest closed path in Minkowski space-time shown
in fig.~1 with parallel segments lying on the light-cone. Under integration
over, say, $x_1$ and $x_2$ along the path $C$, the renormalized Green functions
$G_{\mu_1\cdots\mu_n}(x_1,x_2,\ldots,x_n)$ may have singularities
at coinciding points ($x_1=x_2$) and light-cone singularities
($(x_1-x_2)^2=0$), which give rise to very specific divergences of $W_C.$

These divergences were considered for a long time \ci{Pol} as an undesirable
byproduct in dealing with the loop equations. This opinion radically
changed after it was found \ci{IR} that those specific
divergences are of the utmost importance in perturbative QCD. It turned
out indeed that there is an intimate relation between their
renormalization and infrared asymptotics of hard processes in perturbative
QCD. For instance, the Wilson loop expectation values calculated along
paths partially lying on the light-cone obey  renormalization group
equations (RG) which coincide with the Altarelli-Parisi-Lipatov and the
Brodsky-Lepage evolution equations \ci{KM,WL}. The ``bremsstrahlung''
function, well known in QED for a long time, is nothing but the cusp
anomalous dimension of the Wilson loop \ci{IR,KR}. The same function,
called velocity dependent anomalous dimension, was rediscovered recently
within the heavy quark effective field theory \ci{Geo}. That is why the
investigation of the renormalization properties of Wilson loops has not a
pure academic interest.

In a previous paper \ci{WL} the calculation of the Wilson loop along the
light-like path of fig.~1 was performed in the second order of perturbation
theory in the Feynman gauge. The dimensional regularization was used and
all divergences were subtracted in the ${\MS}-$scheme. It was shown that the
renormalized Wilson loop, being an even dimensionless function of the scalar
product $(n\cdot n^\star)$, of the renormalization point $\mu$ and of
the coupling constant, has a form
$$
W^R_C=W^R_C(\rho,g),
\qqqquad
\rho^2=(n\cdot n^\star)\mu^2,
$$
and obeys the following RG equation \ci{WL,RG}:
\be
\left(\rho\frac{\partial}{\partial\rho}+\beta(g)\frac{\partial}{\partial g}
\right) \log W^R_C(\rho,g)=
-2\Gamma_{cusp}(g)(\log(\rho^2+i0)+\log(-\rho^2+i0))
- \Gamma(g),
\lab{RG}
\ee
$n$ and $n^\star$ being the two light-like vectors
$$
n_\mu=(T,0,0,-T), \qquad n^\star_\mu=(L,0,0,L), \qquad
(n\cdot n)=(n^\star\cdot n^\star)=0, \qquad (n\cdot n^\star)=2LT.
$$
with $L,$ $T > 0.$
The unusual property here is that the r.h.s.\ being considered
as anomalous dimension depends on the renormalization point. It implies
that $W_C$ does not renormalize multiplicatively.

Equation \re{RG} contains
two gauge invariant anomalous dimensions: $\Gamma_{cusp}(g)$ and
$\Gamma(g).$ The first one is related to the asymptotic behaviour of
the so-called cusp anomalous dimension and has very interesting
properties. In particular, $\Gamma_{cusp}(g)$ does not depend on the
form of the integration path $C$ and is equal at two-loop order to
\ci{IR,RG}
\be
\Gamma_{cusp}(g)=\alpi C_F+\left(\alpi\right)^2
C_F\left(C_A\left(\frac{67}{36}-\frac{\pi^2}{12}\right)-N_f\frac5{18}\right).
\lab{cusp}
\ee
with $\as = \frac{g^2}{4\pi}$.
Here, $C_F=\frac{N^2-1}{2N}$ and $C_A=N$ are color factors of the gauge
group $SU(N)$.
The anomalous dimension $\Gamma(g)$ does depend on $C$ and it was found at
two-loop order for the path in fig.~1 to be \ci{WL}
\be
\Gamma(g)=-\ccc{2}C_F
          \lr{\lr{7\zeta(3)-\frac{202}{27}+\frac{11}{36}\pi^2}C_A
           +\lr{\frac{28}{27}-\frac{\pi^2}{18}}N_f}.
\lab{gamma}
\ee
The RG equation \re{RG} was checked in Feynman gauge and gauge
invariance implies that it should be fulfilled in any gauge with the same
anomalous dimensions $\Gamma_{cusp}(g)$ and $\Gamma(g).$ This is one of
the statements we are going to verify.

In the present paper we calculate the Wilson loop expectation value
along path of fig.~1 in the light-like axial (LLA) gauge:
\be
n^\mu A_\mu^a(x)=0
\lab{gauge}
\ee
with the gauge fixing vector along one of the sides of the integration path
at the second order of perturbation theory using dimensional
regularization with $D=4-2\varepsilon $.

It is well-known that the free propagator of the gauge field
is more singular in axial gauges owing to the presence of additional
``spurious" singularities. In the LLA gauge we are considering,
a prescription, the Mandelstam-Leibbrandt (ML)  causal prescription
\ci{pres}, has been proposed to consistently \ci{cons} define it as
\be
G_{\mu\nu}(k)=-\frac{i}{k^2+i0}\left(g_{\mu\nu}-\frac{k_\mu n_\nu+k_\nu
n_\mu}{(kn)+i0\cdot(kn^\star)}\right).
\lab{ML}
\ee
We stress that the $+i0$ prescription only specifies how the integration
contour is to be distorted near the pole. In particular it is irrelevant
in the expression
$$\frac{(nk)}{(nk)+i0\cdot(n^\star k)}= 1,$$
always in the sense of the
theory of distributions; one can also prove that Mandelstam's
proposal does indeed coincide with Leibbrandt's one, namely
$$
\frac{1}{(nk)+i0\cdot(n^\star k)}= \frac{(n^\star k)}{(nk)(n^\star k)+i0}.
$$
The vector propagator contains the famous
$(kn)=0$ singularity and, as a consequence, individual Feynman diagrams
contributing to $W_C$ in LLA gauge may contain additional
poles at $D-4.$  These additional poles should be compensated in
the sum of all diagrams contributing to $W_C$,
in order to ensure its gauge invariance.
The mechanism of this compensation is not trivial and
very sensitive to the prescription one uses to define the propagator
\re{ML}. Our purpose it to show that we shall indeed find full consistency
using the ML prescription, at least up to the second order in
the loop expansion.

Calculating Wilson loop along the path of fig.~1 in the LLA
gauge, we will meet three different kinds of divergences: specific light-cone
singularities of Wilson loops, singularities which are peculiar of
the light-cone gauge we are using and conventional divergences of the
Yang-Mills theory. Since the first two of them originate from the same
phase space region of gluon
momenta, \ie gluons propagating along the light-cone, an interplay is
{\it a priori\/} possible destroying their correct compensation. It would
mean that the LLA gauge with the ML prescription is sick.

This doubt was raised in a recent paper \ci{AT}. The leading
singularities of the Wilson loop of fig.~1 were calculated using the
LLA gauge with the ML prescription in the second order of
perturbation theory and the following expression was obtained
$$
W_C\sim \exp\{B \alpha_s^2 C_FC_A(D-4)^{-4}+\CO((D-4)^{-3})\},
$$
$B$ being a suitable constant, which turned out to be different from
the Feynman gauge result: $B=0$.

In what follows we present a complete two-loop calculation of the
same dimensionally regularized Wilson loop. As dimensional
regularization does not spoil gauge invariance, we shall verify that
the expression we obtain in the LLA gauge for the
unrenormalized (but dimensionally regularized) Wilson loop {\it exactly
coincides} with the analogous quantity evaluated in Feynman gauge up
to terms vanishing when $D = 4$, \ie not only leading and non-leading
logarithmic terms but up to finite ones. This is a complete check of gauge
invariance and thereby of the correctness of the ML prescription,
which is crucial to this goal.

Then, as a second step, we shall renormalize Wilson loop expectation
value in the $\MS-$scheme according to the theory of renormalization
in LLA gauge with ML prescription developed in refs.\ci{gen}. We show
that it indeed obeys the RG equation \re{RG} with the same anomalous
dimensions \re{cusp} and \re{gamma} as in covariant gauges.

The paper is organized as follows. In Sect.~2 the properties of free gluon
propagator are described and the full set of non vanishing Feynman diagrams
contributing to $W_C$ is identified. In Sect.~3 we determine $W_C$ at the
first order of perturbation theory. Second
order Feynman diagrams are evaluated in Sect.~4 and  summed to get the final
expression for the dimensionally regularized Wilson loop, recovering
the Feynman gauge result. In Sect.~5 the Wilson loop
renormalization, which presents non trivial features in LLA gauge,
is performed in the $\MS-$scheme and a check is made of the
anomalous dimensions against the corresponding ones in Feynman gauge,
thereby supporting the renormalization procedure in LLA gauge developed
in refs.\ci{gen}. Sect.~6 contains concluding remarks. Some
technical details of our calculations are presented in the Appendices.

\sect{Free gluon propagator}

To calculate the vacuum average Wilson loop we use the definition \re{def}.
The
integration path $C$ is shown in fig.~1. It lies on the light-cone and is
formed by two light-like vectors $n$ and $n^\star$ we have already
introduced. As a consequence, in the LLA gauge \re{gauge}
gluons cannot be attached to any segment
of the path $C$ going along the gauge fixing vector. We parameterize the
two remaining segments of the path $C$ as follows:
\be
x_1(t)=n^\star t, \quad t\in [0,1];
\qqqquad
x_2(s)=n+n^\star s, \quad s\in [1,0].
\lab{path}
\ee
Note that parameter $s$ runs from $1$ to $0$ in order to provide
the correct orientation of the integration path.

There are a lot of Feynman diagrams contributing to the Wilson loop
expectation value at the second order of perturbation theory. However,
the number of diagrams one has to evaluate can be drastically reduced
using the non-Abelian exponentiation theorem \ci{NET}.
According to this theorem
\be
W_C\equiv 1+\sum_{n=1}^\infty W^{(n)}
   =\exp\sum_{n=1}^\infty w^{(n)},
\lab{ET}
\ee
where $w^{(n)}$ is given by the contribution of $W^{(n)}$ with the
maximal non-Abelian color factor to the $n-$th order of perturbation theory
which is equal to $C_F$ for $n=1$ and $C_FC_A$ for $n=2.$ The two-loop
diagrams containing color factors $C_F$ and $C_FC_A$  are shown in
figs.~2, 3 and 4 where we omitted symmetric diagrams. It turned out
that, owing to the properties of the gluon propagator with the ML
prescription, many of these diagrams give vanishing contributions.

The peculiar features of the free gluon propagator, when using the
ML prescription for the ``spurious'' singularity, have been
discussed in \ci{prop}. In particular it has been shown that the propagator
behaves in the coordinate representation as a tempered distribution at
variance with the expression one would obtain adopting the Cauchy
principal value
prescription. In this last case additional {\IR} singularities would plague
the component $n^\star_\mu n^\star_\nu G^{\mu\nu}(x)$.

The explicit expressions for the propagator with the ML prescription
in the coordinate representation are
$$
G_{\alpha\beta}(x)=\delta_{\alpha\beta}G(x),\qquad \alpha=1,2 ,
$$
with the causal scalar distribution
$$
G(x)=-\frac1{4\pi^2}\frac1{x^2-i0},
$$
and
$$
G_{-\alpha}(x)=\frac{\partial}{\partial x^\alpha}\int_0^{x^+}
            d \xi G(\xi,x^-,x_T),
\qquad
G_{--}(x)=2\frac{\partial}{\partial x^-}\int_0^{x^+}
            d \xi G(\xi,x^-,x_T),
\qquad
G_{+\mu}(x)=0.
$$
We have used here the light-cone variables $x_-,$ $x_+$ and $x_T$ for
the components of vector $x.$ Their definitions and some useful identities
can be found in Appendix A   .

The origin of this reasonable behavior from a mathematical point of view
can be traced back to the causal nature of the ML prescription in which
the ``spurious'' pole complies with the Feynman poles position in the
complex energy plane so that no pinches occur under Wick's rotation of
the integration contour. From
the physical point of view it has been shown \ci{ghost} that a ``longitudinal''
ghost enter the theory. This ghost, while decoupling in all physical
quantities, is responsible for the mild {\IR} behavior of the gluon
propagator.
Nevertheless, since the propagator is a generalized function,
$local$ limits may not exist.
Indeed, using the expression for the $G_{--}(x)$ component one
immediately finds
that the function
$$
G(x)=n^\star_\mu n^\star_\nu G^{\mu\nu}(x)
$$
is divergent for $x_T=0.$

Then
one uses the dimensional regularization to define it at $x_T=0,$ as
\be
G(x_+,x_T=0,x_-)
\equiv
n^\star_\mu n^\star_\nu G^{\mu\nu}(x_+,x_T=0,x_-)
    =\frac{2\pi^{-D/2}\Gamma(D/2)}{4-D}
\frac{(xn^\star)^2}{(-x^2+i0)^{D/2}}.
\lab{spur}
\ee
Of course, it is the ``spurious''
$(kn)=0$ singularity which is responsible for the appearance of a pole in
\re{spur} for $D=4$.%
\footnote{In the Feynman gauge, the propagator of the gauge field is equal to
          $-\frac{g_{\mu\nu}}{4\pi^{D/2}}\Gamma(D/2-1)(-x^2+i0)^{1-D/2}$
          and is regular for $D=4.$}

It is now straightforward to realize that one- and two-loop diagrams
with all gluons attached to the same contour side (see e.g.
figs.~2a and 3a,b) give zero. Consider first the contribution of the
diagram of fig.~2a:
$$
(ig)^2C_F\mu^{4-D}\int_0^1 ds\int_0^s dt
n^\star_\mu n^\star_\nu G^{\mu\nu}(n^\star(s-t)).
$$
Using the definition \re{spur} we find that it vanishes as
\be
n^\star_\mu n^\star_\nu
G^{\mu\nu}(n^\star(s-t))\sim (n^\star n^\star)^{2-D/2}=0
\lab{zero}
\ee
for $\Re D < 4.$ The diagrams of fig.~3a and 3c also vanish since the same
function enters the corresponding path integrals.
We stress that the property \re{zero} is a peculiar consequence of the
ML prescription. To show this, let us perform again the evaluation of
the diagram of fig.~2a in the {\it momentum\/} representation:
$$
(ig)^2C_F\mu^{4-D} \int_0^1ds\int_0^s dt
\int\frac{dk_+dk_-dk_T}{(2\pi)^D}\e^{-ik_-L\sqrt{2}(s-t)}
\frac{2i}{(2k_+k_--k_T^2+i0)}\frac{k_-}{(k_++i0k_-)}.
$$
It is important to note that $k_+$ does not enter the exponent when the
gluon is attached to the same contour side. As a consequence,
the integral over the $k_+$ component is zero because with the ML
prescription poles of Feynman and ``spurious'' denominators lie
on the same side of the real $k_+$ axis. The same arguments can be
applied to show that the diagram of fig.~3b gives zero. In this case
we have an integral over two momenta which vanishes after one closes
the integration contours over ``+'' components of both momenta without
encountering any pole.

\sect{One-loop calculation}

In the lowest order of perturbation theory, Wilson loop \re{def}
has a single nonvanishing contribution coming from the
diagram of fig.~2b
$$
W_C^{(1)}=(ig)^2C_F\mu^{4-D}\int_0^1 ds\int_1^0 dt
          n^\star_\mu n^\star_\nu G^{\mu\nu}(x_2(t)-x_1(s)),
$$
where the functions $x_1(s)$ and $x_2(t)$ are defined in \re{path}. Notice,
that the vector $x_1(s)-x_2(t)$ lies in the plane of the Wilson loop and
one can use expression \re{spur} for the propagator in Minkowski space-time.
After substitutions of \re{spur} and \re{path} one gets
$$
W_C^{(1)}=\frac{g^2}{\pi^{D/2}}C_F\mu^{4-D}
          \frac{2\Gamma(D/2)}{4-D}(nn^\star)^2
          \int_0^1 ds\int_0^1 dt [2(nn^\star)(s-t)+i0]^{-D/2} .
$$
The singularity of the integrand for $s=t$ comes from the
propagation of the gluon in fig.~2b along the light-cone from point
$n^\star s$ to $n+n^\star t$. After an integration which is trivial
in the region $\Re D < 2$ and a subsequent analytic continuation up
to $\Re D < 4 ,$ one finds
\be
W_C^{(1)}=-\frac{g^2}{(2\pi)^{D/2}}C_F\frac{4\Gamma(D/2-1)}{(4-D)^2}
\left[ (-\rho^2+i0)^{2-D/2}+(\rho^2+i0)^{2-D/2} \right].
\lab{res:1}
\ee

We have obtained the same result in the
{\it momentum\/} representation, \ie integrating first over
the contour and performing the Fourier transform afterwards.
Although calculations are more involved in this way, they can be
done directly in the strip $3 < \Re D < 4.$

We conclude that in the first order of perturbation theory the Wilson
loop expectation value contains a double pole. This expression
coincides exactly with the analogous expression in the Feynman gauge.

\sect{Two-loop calculation}

Calculating the Wilson loop in the second order of perturbation theory we
use non-Abelian exponentiation and restrict ourself only to diagrams
containing the non-Abelian color factor $C_FC_A$. As was shown in Sect.~2,
among these diagrams only those shown in fig.~4 give non vanishing
contributions whereas diagrams of figs.~3a,b,c do vanish.

\subsection{Crossed diagram}

The contribution of the ``crossed'' diagram of fig.~4a to $W_C$ contains
the integration of two propagators over the path:
$$
W^{(2)}_{crossed}=(ig)^4C_F(C_F-C_A/2)
\int_0^1 ds_1 \int_0^{s_1} ds_2 \int_1^0 dt_1 \int_1^{t_1} dt_2
G(n+n^\star(t_1-s_1)) G(n+n^\star(t_2-s_2)).
$$
After substitution of \re{spur} $W^{(2)}_{crossed}$ acquires a double
spurious pole
before the integration over $s-$ and $t-$parameters which are ordered along
the path as shown in fig.~4a
\baa
W^{(2)}_{crossed}&=&\frac{g^4}{\pi^D}C_F(C_F-C_A/2)\mu^{8-2D}
                   \frac{4\Gamma^2(D/2)}{(4-D)^2}
                   (nn^\star)^4
\\  \noalign{\vskip 0.2cm}
&\times&
\int_0^1 ds_1 \int_0^{s_1} ds_2 \int_0^1 dt_1 \int_{t_1}^1 dt_2
\left[(2(nn^\star)(s_1-t_1)+i0)(2(nn^\star)(s_2-t_2)+i0)\right]^{-D/2}.
\eaa
As in the previous case, we expect to get additional singularities
from the propagation of both gluons along the light-cone which
correspond to the following values of parameters: $s_1=s_2=t_1=t_2.$
Indeed, a careful integration for $ \Re D < 2$ and therefrom an analytic
continuation up to $ \Re D < 4$ leads to
\ba
W^{(2)}_{crossed}=&&
\frac{g^4}{(2\pi)^D} C_F(C_F-C_A/2) \frac{4\Gamma^2(D/2-1)}{(4-D)^4}
\left\{
\frac{D-2}{D-3} \left[(-\rho^2+i0)^{4-D}+(\rho^2+i0)^{4-D} \right]
\right.\nonumber \\  \noalign{\vskip 0.2cm} &&\left.
+4\left(1-2\frac{\Gamma^2(3-D/2)}{\Gamma(5-D)}\right)
 \left[(-\rho^2+i0)(\rho^2+i0)\right]^{2-D/2}
\right\}.
\lab{res:2a}
\ea
Again the result has been checked in the {\it momentum\/} representation,
where the calculation can be directly performed in the strip $3 < \Re D
<4.$

$W^{(2)}_{crossed}$ has a {\it third\/} pole at $D=4.$ Thus, the leading
singularity of $W^{(2)}_{crossed}$ is formed by a double pole arising
from the ``spurious" contributions and
only one single light-cone singularity. At the same time an analogous crossed
diagram in the Feynman gauge gives rise to a fourth-order pole which
however cancels in the sum of all diagrams.
It means that the Feynman gauge is more singular than LLA gauge as far
as light-cone singularities are concerned.
This property is well known in perturbative QCD.
As we shall show below, all the diagrams of fig.~4 have no fourth pole at all.


\subsection{Self-energy diagram}

The contribution of the diagram of fig.~4b to the Wilson loop reads:
\be
W^{(2)}_{self}=(ig)^2C_F\mu^{4-D}\int_0^1 ds \int_1^0 dt G_1(n+n^\star(t-s)),
\lab{W_self}
\ee
where in the momentum representation the function $G_1(k)$ is defined
as
\be
G_1(k)=\frac12 n^\star_\mu
       G^{\mu\nu}(k)\Pi_{\nu\rho}(k)G^{\rho\lambda}(k)
       n^\star_\lambda.
\lab{self}
\ee
Here $G_{\mu\nu}(k)$ is the free gluon propagator \re{ML} and
$\Pi_{\nu\rho}(k)$ is the one-loop gluon self-energy operator in the
LLA gauge. There are no Faddeev-Popov ghosts and
$\Pi_{\nu\rho}(k)$ gets
contribution only from the gluon loop.%
\footnote{We do not explicitly consider the contribution of the quark
          loop, which does not entail any new feature with respect to
          the one in Feynman gauge \ci{gen}.}

The calculation of the one-loop self-energy $\Pi_{\nu\rho}(k)$
in the LLA gauge is a very cumbersome problem.
As was shown in \ci{trans},
the self-energy $\Pi_{\nu\rho}(k)$ is transverse and it is decomposed
into seven independent tensor structures some of which depending on
the gauge fixing vectors $n$ and $n^\star.$ Moreover the
contributions to the self-energy of the tensors involving $n_\mu$ and
$n^\star_\mu$  exhibit non polynomial residues at $D=4$,
whereas the residue of the contribution
of the ``physical''tensor
($g_{\nu\rho}k^2-k_\nu k_\rho$) gives rise in the LLA
gauge to the one-loop $\beta-$function, since $Z_1=Z_3$ in this gauge
as is well known.

Notice that, after substitution of the self-energy into \re{self}, all
tensors do contribute to the function $G_1(k)$ and the residue at the
pole differs from the $\beta-$function.

The one-loop self-energy was calculated in \ci{trans} up to
$\CO((D-4)^1)-$terms. It is important to recognize
that these terms, which vanish formally for $D=4$, might become divergent
after integration over the path in \re{W_self} due to light-cone
singularities.
This is the reason why one is obliged to first calculate the self-energy for an
arbitrary $D,$ then to
integrate it over the path and only eventually perform a Laurent expansion at
$D=4$. We have performed this
calculation  in the momentum representation and some details are
given in Appendix B  . We present here only the final expression for the
function \re{self}
$$ G_1 (k) = -\frac{4 (n n^\star)^2}{k^2 [nk]} {\cal I}_1
+ 4\left\{ \frac{(n n^\star)^2}{[nk]} - 4 \frac{(n^\star k)(n
n^\star)}{k^2} \right\} {\cal I}_2  - 2\frac{3D -2}{D-1}
\frac{(n^\star k)(n
n^\star)}{k^2 [nk]}{\cal I}_3 - 8\frac{(n n^\star)}{k^2} {\cal I}_4
$$
where the integrals ${\cal I}_1$ ... ${\cal I}_4$ are listed in Appendix
B.
We have checked that $G_1 (k)$, after
Laurent expansion around $D=4$, coincides with the expression \re{self}
calculated starting from the self-energy $\Pi_{\nu\rho}(k)$ found in
\ci{trans}.

By performing the Fourier transform, we obtain the function $G_1$ in
 the {\it coordinate\/} representation
\ba
G_1(x)&=&-\frac{g^2}{8\pi^D}C_A\mu^{4-D}\frac{(xn^\star)^2}{(-x^2+i0)^{D-2}}
   \left\{
       \frac8{(4-D)^3}\left(\frac{\Gamma^2(3-D/2)\Gamma(D-3)}{\Gamma(5-D)}
                            -\Gamma^2(D/2-1)\right)
   \right. \nonumber \\  \noalign{\vskip 0.2cm} && \left.
      +\frac{\Gamma^2(D/2-1)}{(4-D)^2}\left(6-\frac{3D-2}{2(D-1)}
                            -\frac{2(D-2)}{D-3}\right)
   \right\},
\lab{self-x}
\ea
which is valid for $x_T=0$.

Inserting \re{self-x} into \re{W_self} and integrating over the path,
in a region $\Re D<2,$ we get the
following contribution from the diagram of fig.~4b to the Wilson loop
\ba
W^{(2)}_{self}&=&\frac{g^4}{(2\pi)^D}C_FC_A
\left[ (-\rho^2+i0)^{4-D}+(\rho^2+i0)^{4-D} \right]
\nonumber \\  \noalign{\vskip 0.2cm} && \times
\left\{
       \frac4{(4-D)^4(D-3)}\left(\frac{\Gamma^2(3-D/2)\Gamma(D-3)}{\Gamma(5-D)}
                            -\Gamma^2(D/2-1)\right)
   \right. \nonumber  \\  \noalign{\vskip 0.2cm} && \left.
      +\frac{\Gamma^2(D/2-1)}{(4-D)^3(D-3)}\left(3-\frac{3D-2}{4(D-1)}
                            -\frac{D-2}{D-3}\right)
   \right\},
\lab{res:2b}
\ea
which can be continued to the ``natural" strip $3 < \Re D <4.$
Again the same result can be recovered directly in this strip by
performing the calculation in the momentum representation.

The expression for $W^{(2)}_{self}$ exhibits
the third pole coming from the double pole of the function
\re{self-x} and the single light-cone singularity of the path integral in
\re{W_self} for $s=t.$

\subsection{Three gluon diagram}

The calculation of the diagram with the three-gluon vertex is the most
cumbersome one. At variance with the previous diagrams where we have integrated
free gluon propagators in the coordinate representation along the path,
here we have to deal with the additional vertex of a self-gluon
interaction. In the coordinate representation it implies integration over
the intermediate point $z.$ Then the contribution of the diagram of fig.~4c
is given by
\ba
W^{(2)}_{3-gluon}&=&(ig)^4\frac{i}2C_FC_A \mu^{8-2D}
\int_1^0 ds_1 \int_0^1 ds_2 \int_0^{s_2} ds_3
\nonumber
\\  \noalign{\vskip 0.2cm} &&
\times\int d^Dz \int \frac{d^Dk}{(2\pi)^D}
\e^{ik_1(x_2(s_1)-z)}\e^{ik_2(x_1(s_2)-z)}\e^{ik_3(x_1(s_3)-z)}
V(k_1,k_2,k_3),
\lab{W_3gluon}
\ea
where the parameters $s_i$ are ordered along the path and gluons are attached
to
the path at points $x_1(s_1),$ $x_2(s_2)$ and $x_2(s_3)$ defined in
\re{path}.
Here $\frac{d^Dk}{(2\pi)^D}\equiv\prod_{i=1,2,3}\frac{d^Dk_i}{(2\pi)^D}$
denotes integration over gluon momentum and
\be
V(k_1,k_2,k_3)\equiv -i\Gamma_{\mu\nu\rho}(k_1,k_2,k_3)
G_{\mu\mu'}(k_1)G_{\nu\nu'}(k_2)G_{\rho\rho'}(k_3)
n^\star_{\mu'}n^\star_{\nu'}n^\star_{\rho'},
\lab{V}
\ee
$\Gamma_{\mu\nu\rho}(k_1,k_2,k_3)$ being the standard three-gluon
vertex expression.
Note that integration over the intermediate point $z$
ensures momentum conservation $k_1+k_2+k_3=0.$

Our strategy for calculating
$W^{(2)}_{3-gluon}$ is the following.
Instead of integrating over $z$, we first work out the expression
for $V(k_1,k_2,k_3)$
in the momentum representation, then perform independent integrations
over the three gluon momenta in \re{W_3gluon} and  integrate
over $z$ at the very end.

After some algebra we get from \re{V} an expression for $V(k_1,k_2,k_3)$
involving four addenda
\be
V(k_1,k_2,k_3) = V_1 + V_2 + V_3 + V_4,
\lab{V1234}
\ee
where the expression for $V_1$ contains only one single ``spurious" denominator
$[kn] \equiv (kn)+i0\cdot (kn^\star)$:
$$
V_1=\frac{2(nn^\star)}{k_1^2k_2^2k_3^2}
\left\{
  (k_1n^\star)\left(
              \frac{(k_3n^\star)}{[k_3n]}-\frac{(k_2n^\star)}{[k_2n]}
              \right)
 +(k_2n^\star)(k_3n^\star)\left(\frac1{[k_2n]}-\frac1{[k_3n]}
              \right)
 +(k_2-k_3)n^\star\frac{(k_1n^\star)}{[k_1n]}
\right\},
$$
the expression for $V_2$ has two ``spurious'' denominators
$$
V_2=\frac{(nn^\star)^2(k_1n^\star)}{[k_2n][k_3n]k_1^2}
    \lr{\frac1{k_2^2}-\frac1{k_3^2}}
   +\frac{(nn^\star)^2}{[k_1n]k_1^2}
    \lr{\frac{(k_3n^\star)}{[k_2n]k_3^2}-\frac{(k_2n^\star)}{[k_3n]k_2^2}}
   +\frac{(nn^\star)^2}{[k_1n]k_2^2k_3^2}
    \lr{\frac{(k_2n^\star)}{[k_3n]}-\frac{(k_3n^\star)}{[k_2n]}}
$$
and $V_3$ and $V_4$ depend only on two  momenta
$$
V_3=\frac{(nn^\star)^2(k_3-k_2)n^\star}
{[k_2n][k_3n]k_2^2k_3^2},
\qquad
V_4=\frac{(nn^\star)^2(k_2-k_1)n^\star}{[k_1n][k_2n]k_1^2k_2^2}
   +\frac{(nn^\star)^2(k_1-k_3)n^\star}{[k_1n][k_3n]k_1^2k_3^2}.
$$
It turns out that, after substitution of $V_4$ into \re{W_3gluon} and
integration
over momenta $k_2$ and $k_3$, we get a vanishing result even before
integration over the parameters $s_i$.

Consider the first term in the
expression $V_4$. Its integration over $k_3$ leads to $z=x_1(s_3)$ and
the resulting expression, being considered as a function of $(k_2n)$, has
singularities coming from two propagators $\frac1{[k_2n]}$ and
$\frac1{k_2^2+i0}$ which lie on the same side of real axis in the
complex $(k_2n)-$plane.%
\footnote{We have used here the identity
          $k^2=2(kn)(kn^\star)/(nn^\star)-k_T^2$ with $k_T=(k_1,k_2).$}
Hence the integral over the $(k_2n)$ component of momentum $k_2$
vanishes. The same argument applies to the second term in $V_4$ and
therefore one concludes that $V_4$ does not contribute to \re{V1234}.

Substituting expressions for
$V_1$, $V_2$ and $V_3$ into \re{W_3gluon} one finds that the result of
the integration over momenta $k_1,$ $k_2$ and $k_3$ can be expressed in terms
of the following three basic functions:
\ba
F_1(x)&=&\int \frac{d^Dk}{(2\pi)^D}\e^{ikx} \frac1{(k^2+i0)[kn]},
\nonumber \\  \noalign{\vskip 0.2cm}
F_2(x)&=&\int \frac{d^Dk}{(2\pi)^D}\e^{ikx} \frac1{(k^2+i0)},
\lab{F} \\  \noalign{\vskip 0.2cm}
F_3(x)&=&\int \frac{d^Dk}{(2\pi)^D}\e^{ikx} \frac1{[kn]}.
\nonumber
\ea

Let us first consider the contribution of the expression $V_3$ into
$W^{(2)}_{3-gluon}$.
Since $V_3$ does not depend on $k_1$, integration over this momentum leads
to a $\delta-$function implying $z=x_2(s_1)=n+n^\star s_1$ and we eventually
get
$$
\frac12g^4C_FC_A\mu^{8-2D}(nn^\star)^2
\int_0^1 ds_1 \int_0^1 ds_2 \int_0^{s_2} ds_3
\lr{\fracs{\partial}{\partial s_2}-\fracs{\partial}{\partial s_3}}
F_1(-n-n^\star(s_1-s_2)) F_1(-n-n^\star(s_1-s_3)).
$$
Using the expression \re{F-lc} for the function $F_1(x)$ with the vector $x$
lying in the
plane of the path $C$ given in Appendix  C, we integrate over the
parameters $s_i$ and obtain the contribution to $W^{(2)}_{3-gluon}$
arising from $V_3$
\ba
-\frac{g^4}{(2\pi)^D}C_FC_A\frac{\Gamma^2(D/2-1)}{(4-D)^4}
&& \left\{ \frac{8\Gamma^2(3-D/2)}{\Gamma(5-D)}
\left[(-\rho^2+i0)(\rho^2+i0)\right]^{2-D/2}
\right.
\nonumber \\  \noalign{\vskip 0.2cm}
&& \left.
-\frac1{D-3}
\left[ (-\rho^2+i0)^{4-D}+(\rho^2+i0)^{4-D} \right]
\right\}.
\lab{V3}
\ea
This expression has a fourth order pole in the dimensional regularization
parameter and contains two different structures as an even function
of $(nn^\star)$.

Calculation of the integrals coming from $V_1$ and $V_2$
is more complicated. We shall only sketch here the evaluation of one of them
taking only two terms from $V_1$ as an example:
$$
\frac{2(k_2n^\star)(k_3n^\star)(nn^\star)}{k_1^2k_2^2k_3^2}
\lr{\frac1{[k_2n]}-\frac1{[k_3n]}}.
$$
Replacing the vertex $V$ by this expression in \re{W_3gluon}, one gets
\ba
J&=&ig^4C_FC_A\mu^{8-2D}(nn^\star)
\int_0^1 ds_1 \int_0^1 ds_2 \int_0^{s_2} ds_3
\lab{J}
\\  \noalign{\vskip 0.2cm}
&&\times\int d^Dz
F_2(n+n^\star s_1-z)
\frac{\partial^2}{\partial s_2 \partial s_3}
(F_1(n^\star s_2-z) F_2(n^\star s_3-z)
-F_2(n^\star s_2-z) F_1(n^\star s_3-z)),
\nonumber
\ea
where the functions $F_i$ are defined in \re{F} and their explicit
expressions can be found in Appendix C. At variance with the previous
case, the integral over the intermediate point $z$ is not trivial and
is given by \re{z}.
After substitution of \re{z} into $J$ we integrate over the parameters
$s_i$ and eventually get:
\ba
J=&&-\frac{g^4}{64\pi^D}C_FC_A\frac{\Gamma(D-3)}{(4-D)^2}
\left[ (-2(nn^\star)+i0)^{4-D}+(2(nn^\star)+i0)^{4-D} \right]
\nonumber
\\  \noalign{\vskip 0.2cm}&&\times
\int_0^1 d\beta
\beta_1^{2-D/2}\beta_3^{D/2-2}\frac{1-\beta_2^{D/2-2}}{1-\beta_2}
\frac{M(\beta_2,\beta_3)}{\beta_2+\beta_3},
\lab{M}
\ea
where $d\beta\equiv d\beta_1d\beta_2d\beta_3
\delta(1-\beta_1-\beta_2-\beta_3)$,  $0\leq\beta_i\leq 1$ and
$$
M(\beta_2,\beta_3)=\beta_3^{4-D}-\beta_2^{4-D}+(\beta_2-\beta_3)
(\beta_2+\beta_3)^{3-D}.
$$
The integral over the parameters $\beta_i$ for an arbitrary $D$
can be expressed in the form of a convergent series, which however
does not sum to elementary functions. For our purposes it is enough
to expand it in powers of $(4-D)$ with
the following result:
\ba
J=&&\frac{g^4}{(2\pi)^D}C_FC_A\frac{\Gamma(D-3)}{4(4-D)^2}
\left[ (-\rho^2+i0)^{4-D}+(\rho^2+i0)^{4-D} \right]
\nonumber
\\  \noalign{\vskip 0.2cm}&&\times
\left\{
(1-\zeta(3))(4-D)+ (\fracs52 -2\zeta(3) + \fracs{\pi^4}{144})(4-D)^2 +{\cal
O}((4-D)^3)
\right\},
\lab{V2_1}
\ea
where $\zeta(z)$ is the Riemann function.

The calculation of the remaining terms entering the expressions for $V_1$
and $V_2$ is analogous. For all of them we get  the final expressions
given in \re{I's}, which contain similar integrals over the parameters
$\beta_i$ with the same function $M(\beta_2,\beta_3)$ defined in \re{M}.

Summing all expressions \re{V3} and \re{I's} we obtain the contribution of
the diagram
of fig.~4c to the Wilson loop, we multiply it by factor 2 to take into account
the symmetric diagram with two gluons attached to the lower contour side and
get
\ba
W^{(2)}_{3-gluon}&=&\frac{g^4}{(2\pi)^D}C_FC_A
\left\{
-\left[(-\rho^2+i0)(\rho^2+i0)\right]^{2-D/2}
 \frac{8\Gamma^2(3-D/2)\Gamma^2(D/2-1)}{(4-D)^4\Gamma(5-D)}
\right.
\nonumber
\\  \noalign{\vskip 0.2cm}  &&
+\left[ (-\rho^2+i0)^{4-D}+(\rho^2+i0)^{4-D} \right]
 \Gamma(D-3)
\nonumber
\\  \noalign{\vskip 0.2cm} &&
 \left[ \frac2{(4-D)^4}
      \left(2\Gamma(D/2-1)\Gamma(3-D/2)
           +\frac1{D-3}\frac{\Gamma^2(D/2-1)}{\Gamma(D-3)}
           -\frac{6-D}{D-2}\frac{\Gamma^2(3-D/2)}{\Gamma(5-D)}
      \right)
\right.
\nonumber
\\  \noalign{\vskip 0.2cm} && \left. \left.
      -\frac1{(4-D)^2}\frac{\pi^2}{12}
      +\frac1{4-D}\frac{2-5\zeta(3)}{4}
      +\frac{\pi^4}{80}+\frac54-\zeta(3)
 \right]
\right\}.
\lab{res:2c}
\ea
This complicated expression is valid up to terms vanishing for
$D=4$.

The contribution coming from diagrams with three gluon vertex has
been computed also in {\it momentum} representation, namely integrating
first over the contour and then over momenta. All integrals in this
case can be directly performed in the strip $3< \Re D <4$. Details of the
calculation are reported in Appendix D. The final result of course fully
coincides with expression \re{res:2c} and provides an independent check.

We have calculated in this section all two-loop diagrams contributing
to the Wilson loop expectation value with non-Abelian color factor
$C_FC_A$. To find two-loop {\it unrenormalized\/} Wilson loop
using non-Abelian exponentiation theorem we sum expressions \re{res:2a},
\re{res:2b} and \re{res:2c}, skip terms with color factor $C_F^2$ in front
and identify the resulting expression with the exponent $w^{(2)}$ in \re{ET}.

Notice that unrenormalized Wilson loop is a finite function of $D$
as long as $D\neq 4$. Trying to give a physical meaning to this
quantity,
we have to take care of poles at $D-4$ and to perform the renormalization
procedure, which eventually will provide us with the two-loop
renormalized Wilson
loop for $D=4$. Renormalization will be performed in the next section.

At this stage we can already settle our first problem, namely
whether the LLA gauge with the ML prescription is sick
in the calculation of the Wilson loop under consideration.
We stress that unrenormalized
Wilson loop for $D\neq 4$ is a well defined gauge invariant quantity.
The same unrenormalized Wilson loop was
calculated in \ci{WL} using Feynman gauge. We take here the opportunity
of correcting a misprint in the expression for $W_f$ in ref.\ci{WL},
where the constant $\frac{11}{360}\pi^2$ should be replaced by
$\frac{11}{360}\pi^4$.

Combining two-loop expressions for unrenormalized
Wilson loop in the axial and Feynman gauges and performing their Laurent
expansions near $D=4$ we find that {\it Wilson loops are identical in
both gauges up to terms vanishing for $D=4$.\/} It means that LLA
gauge with the ML prescription works perfectly, at least up to two-loop order.

\sect{Wilson loop in the light-like axial gauge}

In the previous section we found the two-loop unrenormalized expressions
for all independent diagrams contributing to the light-like Wilson loop.
To give a  meaning to these expressions for $D=4$ we have to define the
renormalization procedure which, when applied to unrenormalized Wilson
loop, removes all divergences including overlapping divergences like
the one of the gluon self-energy in fig.~4b.

Calculating diagrams we have met
three different kinds of divergences: specific light-cone divergences
of the Wilson loops, spurious divergences of the LLA gauge and
conventional divergences of the Yang-Mills theory. All of them have
manifested themselves as poles in the dimensional regularization
parameter.

The problem with the renormalization of the first kind of singularities
has to do with to the fact that Wilson loop is a non-local functional
of the gauge potentials. As was shown in ref.\ci{Pol}, had the integration
path neither cusps nor light-like segments, it would be enough to renormalize
Green functions and coupling constant to remove all divergences
from $W_C$.
But since the integration path under consideration has both kinds of
peculiarities, one should first renormalize the gauge potentials and
coupling constant and then subtract the remaining specific light-cone
and cusp divergences .

The general theory of renormalization in LLA gauge has been developed
in refs.\ci{gen}. In particular it has been shown there that the vector
potential $A_{\mu}(x)$ renormalizes at all orders in the loop expansion
with {\it two} renormalization
constants $Z_3$ and $\tilde Z_3$
\be
A_{\mu}(x)= Z_3^{1/2} [A_{\mu}^R(x) - (1- \tilde Z_3^{-1}) n_{\mu}
\Omega^R(x)],
\ee
with the non-local gauge dependent quantities $\Omega(x)$ being defined in
terms of the covariant derivative
${\cal D}_\mu^{ab}=\delta^{ab}\partial_\mu + gf^{abc}A_\mu^c$
in the adjoint representation and of the
field tensor as
\be
\Omega^a(x)=[(n{\cal D})^{-1}]^{ab} (n n^\star)^{-1}
n_\mu n^\star_\nu F^{\mu \nu,b}(x).
\ee
The operator $(n{\cal D})^{-1}$ is understood as a series expansion in the
coupling constant with boundary conditions giving rise to the ML
prescription. It is immediate to realize that renormalization does not
change the gauge condition.

To our subsequent calculations, only
the effective part of the zero-th order expansion in the coupling constant
of eq. (5.2)
$$
\Omega^a(x) = \frac{n^\star A^a(x)}{(n^\star n)}
$$
will be relevant.

The coupling constant $g$ is renormalized as $Z_3^{-1/2} g^R$
as $Z_1=Z_3$ in this gauge. As a consequence the Wilson loop gets
the expression in terms of renormalized potentials
\be
W_C=\frac1N\vev{0|\ \Tr\CT\CP\exp\left(ig^R \oint_C dx^\mu [A_{\mu}^R(x) -
(1- \tilde Z_3^{-1}) n_{\mu} \Omega^R(x)] \right)|0}.
\ee
The renormalization constants $Z_3$ and $\tilde Z_3$ at order $g^2$
are given by \ci{Z}
\ba
       Z_3 &=& 1+ \frac{11g^2}{24\pi^2}\frac{C_A}{4-D},
\\
\tilde Z_3 &=& 1 + \frac{g^2}{4 \pi^2} \frac{C_A}{4-D}.
\ea
The extra-term in the Wilson loop (5.3) gives rise at $\CO (g^4)$ only to
a contribution coming from the cross-product, which turns out to be
equivalent to the following correction to the gluon propagator
\be
\delta G_{\mu\nu}(x)=\frac{g^2}{2\pi^2}\frac{C_A}{4-D}G_{\mu\nu}(x).
\ee
In turn, the divergent part at $D=4$ of the gluon propagator due to
the self-energy insertion of fig.4b can easily be extracted from
\ci{trans}. It originates from the following two contributions of the
self-energy tensor
\be
\Pi_{\nu\rho}(k)= \frac{i g^2}{8 \pi^2} C_A
\left\{ {11 \over 3}
\frac1{4-D}(k^2 g_{\nu\rho} - k_\nu k_\rho) + \frac4{4-D}
\frac{(nk)}{(n n^\star)} (k_\nu n^\star_\rho + k_\rho n^\star_\nu)
\right\}.
\ee
One can easily check that the contribution from the second tensor is
{\it exactly} compensated by the correction in the expression of the
Wilson loop due to the presence of $\Omega$. Therefore only the term
which is proportional to the
$\beta-$function is to be subtracted and the resulting path integral
coincides (up to factor
$-\frac{11g^2}{24\pi^2} C_A\frac1{4-D}$)
with one-loop expression \re{res:1}
\be
W^{(2)}_{c.-t.}= \frac{g^4}{8\pi^D}\frac{11}3
C_FC_A\frac{\Gamma(D/2-1)}{(4-D)^3}
\left[ (-2(nn^\star)+i0)^{2-D/2}+(2(nn^\star)+i0)^{2-D/2} \right].
\ee
Once this expression has been subtracted from the sum of contributions
to the Wilson loop, turning the coupling constant $g$ into $g^R$,
not all divergent parts at $D=4$ are eliminated.
The remaining divergences are just light-cone divergences of the
Wilson loop, which we also have to subtract taking care of overlapping
singularities. Here we meet some simplifications in the
structure of light-cone singularities in LLA gauge.
It is indeed well known from perturbative QCD that diagrams like
the one of fig.~4, having non-Abelian color factor $C_FC_A$ and either
being nonplanar (see fig.~4a) or containing gluon self-interaction, do
not exhibit overlapping light-cone (or collinear) divergences.
It means that, in order to renormalize those diagrams, one can simply
subtract all poles remaining after renormalization of
Green functions. Then, using the non-Abelian
exponentiation theorem, one realizes that, in order to renormalize
all light-cone singularities of the Wilson loop $W$, it is enough
to subtract poles from the exponent $w^{(n)}$, which does not contain any
overlapping light-cone singularity.

After having subtracted those remaining poles in the $\MS-$scheme
according to this procedure, we verify that the resulting renormalized
expression satisfies the RG equation \re{RG} with the same anomalous
dimensions of eqs. \re{cusp} and \re{gamma}, providing a further
test of gauge invariance and, at the same time, of the way in which
renormalization operates in the LLA gauge.

\sect{Conclusions}

We end by summarizing what we have achieved in previous sections.
We have first evaluated in perturbation theory at $\CO(g^4)$
a light-like Wilson loop for the Yang-Mills theory in LLA gauge
with the ML prescription for the vector propagator. All singularities,
the ones related to the ultraviolet behaviour of the theory, the
ones peculiar of the contour we have chosen (light-like lines
and cusps) and the ones related to the gauge choice, have been
dimensionally regularized and manifest themselves as poles at $D=4$.

As dimensional regularization does not spoil gauge invariance,
we have checked that our result does indeed completely coincide
with the analogous one obtained in ref.\ci{WL} for the same contour
in Feynman gauge. In this way a  complete test at $\CO(g^4)$
of the correctness of the ML prescription in LLA gauge has
been achieved.

Then we have renormalized our result in the $\MS-$scheme. In so doing
we explicitly extend the general prescriptions given in refs.\ci{gen}
to a non-local operator (the Wilson loop). We again recover in a
quite non trivial way the Feynman gauge findings; in particular
both cusp and anomalous dimensions, which obey the
RG equation and are relevant to the physics of  soft
radiation, do indeed respectively coincide with their Feynman gauge
counterparts, providing a further test in favour of general
procedures concerning canonical quantization and renormalization
in LLA gauge as well as of its usefulness in practical calculations.

\noindent
\bigskip\par{\Large {\bf Acknowledgements}}\par\bigskip\par

Two of us (I.A.K. and G.P.K.) would like to thank G.Marchesini for
numerous and useful discussions.

\sect{Appendices}

{\Large {\bf Appendix A. Light-cone variables}}
\renewcommand\theequation{A.\arabic{equation}}
\setcounter{equation} 0

\bigskip
\noindent
For an arbitrary vector $x=(x_0,x_1,x_2,x_3)$ light-cone variables
are defined as
$$
x_+=\frac1{\sqrt{2}}(x_0+x_3),
\qquad
x_-=\frac1{\sqrt{2}}(x_0-x_3),
\qquad
x_T=(x_1,x_2)
$$
The following identities are fulfilled for two 4-dimensional vectors
$x$ and $y$:
$$
x_\pm=x^\mp,
\quad
x^2=2x_+x_--x^2_T,
\quad
(xy)=x_+y_-+x_-y_+-(x_T\cdot y_T),
\quad
d^4x=dx_+dx_-d^2x_T
$$
For gauge fixing vectors $n$ and $n^\star$ we have
$$
n_+=0, \quad n_-=\sqrt{2}T, \quad n_T=0,
\qquad
n_+^\star=\sqrt{2}L, \quad n_-^\star=0, \quad n_T^\star=0.
$$

\bigskip
\noindent
{\Large
{\bf Appendix B. Useful integrals in the evaluation of the self-energy
diagram}
}
\renewcommand\theequation{B.\arabic{equation}}
\setcounter{equation} 0
\bigskip

\noindent
We list here the
 integrals used in the expression  $G_1 (p)$.  We follow the conventions
given in Appendix A for light-cone coordinates.
In the defining integrals, ML prescription for the spurious poles will
always be understood.
\ba
{\cal I}_1 (p) &=& \int \frac{d^D k}{(2\pi)^D} \frac{1}{k^2 [p_++k_+]}
\nonumber \\ \noalign{\vskip 0.2cm}
 &=& -2ip_- \frac{\Gamma(1-D/2)}{(4\pi)^{D/2}} (-2p_+p_-
-i0)^{D/2 -2}\\ \noalign{\vskip 0.2cm}
 {\cal I}_2 (p) &=& \int \frac{d^D k}{(2\pi)^D} \frac{1}{k^2 (p+k)^2
[k_+]}
\nonumber \\ \noalign{\vskip 0.2cm}
 &=& \frac{i}{(4\pi)^{D/2}} \frac{(-p^2 -i0)^{D/2 -2}}{[p_+]}
\left\{ \frac{\Gamma(3-D/2)}{(2-D/2)^2} \left[\frac{\Gamma^2 (D/2
-1)}{\Gamma(D-3)} -1 \right] \right.\nonumber  \\  \noalign{\vskip 0.2cm} &&
+ \left.\frac{1}{2\pi
i}\int_{\eta-i\infty}^{\eta+i\infty} ds \, \frac{\Gamma(2-D/2
+s) \Gamma (-s)}{D/2 -2 +s} \left(\frac{p^2_T}{p^2
+i0}\right)^s\right\}\\  \noalign{\vskip 0.2cm}
 {\cal I}_3 (p) &=& \int \frac{d^D k}{(2\pi)^D} \frac{1}{k^2 (p+k)^2}
\nonumber \\ \noalign{\vskip 0.2cm}
 &=& \frac{i}{(4\pi)^{D/2} }
\frac{\Gamma (2-D/2) \Gamma^2 (D/2 -1)}{\Gamma (D-2)} (-p^2 -i0)^{D/2
-2} \\ \noalign{\vskip 0.2cm}
{\cal I}_4 (p) &=& \int \frac{d^D k}{(2\pi)^D} \frac{k_-}{k^2 (p+k)^2
[k_+]}
\nonumber \\ \noalign{\vskip 0.2cm}
 &=& \frac{i}{(4\pi)^{D/2} }\frac{(-p^2 -i0)^{D/2-2}}{[p_+]}
\left\{ p_- \Gamma (2-D/2)\left[\frac{\Gamma^2 (D/2 -1)}{\Gamma(D-2)}
-\frac{1}{D/2 -1} \right] \right. \nonumber \\ \noalign{\vskip 0.2cm}
&&\left. + \frac{(p^2 +i0)}{2[p_+]} \Gamma (1-D/2)
\left[\frac{\Gamma (D/2 -2) \Gamma (D/2)}{\Gamma (D-2)}+
\frac{D/2 -3}{D/2 -2} \right]\right\} \nonumber \\ \noalign{\vskip 0.2cm}
&&-\frac{i}{(4\pi)^{D/2} }\frac{(-p^2 -i0)^{D/2-1}}{4[p_+]^2}
\int_{-\eta-i\infty}^{-\eta+i\infty} ds\, \frac{\Gamma (3+s -D/2)}{(s+D/2)
\Gamma (3+s) \sin \pi s}
\left( \frac{p^2_T}{p^2 +i0} \right)^{s+2}
\ea
In Eqs. (B.2) and (B.4), $\eta \in (0,1)$.

\bigskip
\noindent
{\Large
{\bf Appendix C. Diagram with three-gluon vertex (coordinate
representation)}
}
\renewcommand\theequation{C.\arabic{equation}}
\setcounter{equation} 0
\bigskip

\noindent Performing calculation of the diagram with a three-gluon vertex
we have introduced three functions defined in \re{F}. Their explicit
expressions are
\ba
F_1(x)&=&\frac{i^{1-D/2}}{4\pi^{D/2}}\frac{(xn^\star)}{(nn^\star)}
         \int_{0}^{\infty}d\alpha\alpha^{D/2-2}
         \int_{0}^{1} d\xi \xi^{1-D/2}
         \e^{ -i\alpha\left(2x_-x_+ -x_T^2/\xi\right)}
\nonumber \\  \noalign{\vskip 0.2cm}
F_2(x)&=&\frac{i^{-D/2}}{4\pi^{D/2}}
         \int_{0}^{\infty}d\alpha\alpha^{D/2-2}\e^{ -i\alpha x^2}
       =-\frac{i}{4\pi^{D/2}}(-x^2+i0)^{1-D/2},
\lab{F's}
\\ \noalign{\vskip 0.2cm}
F_3(x)&=&\frac{i}{\pi}\frac{(xn^\star)}{(nn^\star)}\delta(x_T)
         \int_{0}^{\infty}\ d\alpha\e^{ -i\alpha x^2}
       =\frac1{\pi}\frac{(xn^\star)}{(nn^\star)}
        \frac{\delta^{(D-2)}(x_T)}{x^2-i0}
\nonumber
\ea
for arbitrary $D-$dimensional vector $x$ in the coordinate space. Although
some integrals over $\alpha-$parameters can be done here we remain them
undone in order to simplify further integration of the product of these
functions over $x.$
Expression for $F_1(x)$ is greatly simplified as the vector $x$ lies
in the plane of the vectors $n$ and $n^\star$:
\be
F_1(x_+,x_T=0,x_-)=\frac1{2\pi^{D/2}}\frac{\Gamma(D/2-1)}{4-D}
\frac{(xn^\star)}{(nn^\star)}(-x^2+i0)^{1-D/2}\,.
\lab{F-lc}
\ee

Using \re{F-lc} we have calculated in \re{V3} the contribution of $V_3-$term.
Calculation
of the contribution due to $V_1$ and $V_2$ terms involves integration of
products of functions \re{F's} over intermediate point $z$. One of these
integrals entering into \re{J} can be calculated using \re{F's} as
\ba
\int d^Dz F_2(n+n^\star s_1-z)
\lr{F_1(n^\star s_2-z)F_2(n^\star s_3-z)
   -F_2(n^\star s_2-z)F_1(n^\star s_3-z)}
\nonumber
\\ \noalign{\vskip 0.2cm}
=\frac{i}{32\pi^D}\frac{\Gamma(D-3)}{4-D}
\int_0^1 d\beta
\beta_1^{2-D/2}\beta_3^{D/2-2}\frac{1-\beta_2^{D/2-2}}{1-\beta_2}
(\Phi(\beta_1,\beta_2,\beta_3)-\Phi(\beta_1,\beta_3,\beta_2))
\lab{z}
\ea
where $d\beta\equiv d\beta_1d\beta_2d\beta_3
\delta(1-\beta_1-\beta_2-\beta_3)$,  $0\leq\beta_i\leq 1$ and the notation
was used for the function
$
\Phi(\beta_1,\beta_2,\beta_3)=
[2(nn^\star)(\sum_{i=1}^3 s_i\beta_i-s_1)+i0]^{3-D}
$
The $\beta-$parameters in \re{z} have appeared after one has performed the
following change of the integration variables:
$\alpha_i=\lambda\beta_i$ ($i=1,2,3$) and integrated over $\lambda$
from $0$ to $\infty$.
Substituting \re{z} into \re{J} and integrating over $s_i-$parameters we
get \re{M}.

Different terms entering into expressions for $V_1$ and $V_2$ are
grouped into 6 items. Contribution of one of them is given by \re{V2_1}.
The
calculation of the remaining 5 terms is analogous. One first rewrites them
as integrals of a product of $F-$functions over $z$, uses relations \re{F's}
to express the result of $z-$integration as integral over
$\alpha-$parameters and then integrates over $s-$parameters taking into
account their ordering along the path. At the end of this procedure
substituting $\alpha_i=\lambda\beta_i$ we get integrals over
$\beta-$parameters analogous to \re{M}. Summarizing, we can write the
contribution of 6 items from $V_1$ and $V_2$ to the diagram of fig.~4c as
\be
\frac{g^4}{64\pi^D}C_FC_A
\left[ (-2(nn^\star)+i0)^{4-D}+(2(nn^\star)+i0)^{4-D} \right]
\frac{\Gamma(D-3)}{(4-D)^2}
(I_1+I_2+I_3+I_4+I_5+I_6)
\lab{I's}
\ee
where $I_1,I_2,I_3$ correspond to three items in $V_1$ and
$I_4,I_5,I_6$ to three items in $V_2$:
\baa
I_1&=&-\int_0^1 d\beta
\beta_1^{2-D/2}\beta_3^{D/2-2}\frac{1-\beta_2^{D/2-2}}{1-\beta_2}
\frac{M(\beta_2,\beta_3)}{\beta_2+\beta_3}\\ \noalign{\vskip 0.2cm} &
= &(1-\zeta(3))(4-D)+ (\fracs52 -2\zeta(3) + \fracs{\pi^4}{144})(4-D)^2 +{\cal
O}((4-D)^3)
\\ \noalign{\vskip 0.2cm}
I_2&=&\int_0^1 d\beta
\beta_1^{1-D/2}\beta_2^{D/2-2}\beta_3^{D/2-3}(1-\beta_1^{2-D/2})
M(\beta_2,\beta_3)
\\ \noalign{\vskip 0.2cm} & = &\frac{4\Gamma(D/2-1)}{(4-D)^2}
    \lr{\Gamma(3-D/2)-\frac{\Gamma(D/2-1)}{2\Gamma(D-2)}}
\\ \noalign{\vskip 0.2cm}
I_3&=&-\int_0^1 d\beta
\beta_1^{2-D/2}\beta_3^{D/2-3}\frac{1-\beta_2^{D/2-2}}{1-\beta_2}
M(\beta_2,\beta_3)
\\ \noalign{\vskip 0.2cm} &=& \fracs{\pi^2}6-\fracs52\zeta(3)(4-D)
+ \fracs{21}{720}\pi^4 (4-D)^2 + {\cal O}((4-D)^3)
\\ \noalign{\vskip 0.2cm}
I_4&=&\int_0^1 d\beta
\beta_1^{3-D}\beta_2^{D/2-3}\beta_3^{D/2-2}
M(\beta_2,\beta_3)
=-\frac{4\Gamma(D/2-1)}{(4-D)^2}
    \lr{\Gamma(3-D/2)-\frac{\Gamma(D/2-1)}{2\Gamma(D-2)}}
\\ \noalign{\vskip 0.2cm}
I_5&=&-\int_0^1 d\beta
\beta_1^{2-D/2}\beta_2^{D/2-3}\beta_3^{-1}
M(\beta_2,\beta_3)
\\ \noalign{\vskip 0.2cm} &=&\frac{2\Gamma ( D/2 -1) \Gamma
(3-D/2)}{(4-D)^2} - \frac{2\Gamma^2 (3-D/2)}{ (4-D)\Gamma
(5-D) }[\psi(3-D/2)-\psi (D/2-1)]
\\ \noalign{\vskip 0.2cm}
I_6&=&\int_0^1 d\beta
\beta_1^{1-D/2}\beta_2^{D/2-2}\beta_3^{-1}
M(\beta_2,\beta_3)
\\ \noalign{\vskip 0.2cm}
&=& \frac{2\Gamma ( D/2 -1) \Gamma
(3-D/2)}{(4-D)^2} + \frac{\Gamma (3-D/2) \Gamma(2-D/2)}{\Gamma(5-D)}
\left[\psi(3-D/2) -\psi(D/2) -\frac{1}{(D/2-1)}\right]
\eaa
where $\psi(z) =d\, \log \Gamma (z)/dz$.
Note that $I_1$ was used in \re{M}.

\bigskip
\noindent
{\Large
{\bf Appendix D. Diagram with three-gluon vertex
\par \noindent (momentum representation)}
}
\renewcommand\theequation{D.\arabic{equation}}
\setcounter{equation} 0
\bigskip

\noindent
Among the various diagrams,
the evaluation of the three-gluon diagram  given in Fig. (4.c) is the
most delicate. In the main text and in the previous Appendix we have
shown how the calculation can be performed in the coordinate
representation. Since this is the only diagram in which
the difference between the computation in  the momentum and coordinate
representation is appreciable, it is instructive to sketch the main lines
of calculation  in the momentum representation.

We start from Eq. (4.4). The trivial integration over $dz$ reproduces
the momentum conservation delta distribution. Then, we can write
 the three-gluon diagram as
\be
W_{3-gluon} = \alpha \sum_{i=1}^3 \int \frac{d^D p \, d^Dq \, d^Dr}{
(2\pi)^{2D}} \frac{{\cal F}_i (p, q, r)}{p^2 q^2
r^2} \CG(p , q, r)\delta(p+q+r)
 \equiv \sum_{i=1}^3 W^{(i)}
\ee
where $\alpha =2ig^4 C_A C_F \mu^{8-2D}$,
\ba
{\cal F}_1 (p,q,r)&=&\frac{2}{[p_+]} \left\{(pq)\frac{r_-}{[r_+]}
-(pr)\frac{q_-}{[q_+]}\right\} +\left\{ \frac{}{} {\rm cyc. \ perm. \ of} \
(p,q,r)\right\}\\ \noalign{\vskip 0.2cm}
{\cal F}_2 (p,q,r)&=&\frac{2p_-}{[p_+]} \{ q_--r_-\} \\ \noalign{\vskip 0.2cm}
{\cal F}_1 (p,q,r)&=&\frac{2q_-}{[q_+]}\{r_- -p_-\} + \frac{2r_-}{[r_+]}
\{p_- - q_-\}
\ea
and $\CG(p,q,r) $ denotes the so called ``geometrical factor'', encoding
all the local geometrical properties of the loop contour
\ba
G(p,q,r)& =& e^{2ip_+T} \int_0^{2L}dx^- e^{ip_-x^-}\int_{2L}^0 dy^-
e^{iq_-y^-} \int_{2L}^{y^-} dz^- e^{ir_-z^-} \nonumber \\ \noalign{\vskip
0.2cm}
&=& \frac{i e^{2ip_+T}}{p_- } \left(e^{2ip_-L} -1 \right)
\left\{ \frac{1}{p_- r_-} \left( e^{-2iLp_- } -1\right) +
\frac{e^{2iLr_-}}{q_-r_-} \left( e^{2iLq_-} -1 \right) \right\}
\ea
In the amplitude factors ${\cal F}_i$, all the spurious singularities
are defined through  ML distributions. $\CG(p,q,r)$, instead, does not
contain any pole, and all the singularities in the $q_-$, $p_-$ and
$r_-$ variables are fictitious.
Since  the ${\cal F}_i (p,q,r)$ are antisymmetric under the
exchange $(q,r) \to (r,q$), the geometrical factor $G$ in the integrand
(D.1) can be more
conveniently rewritten by performing the change of variables
$(q,r) \to (r,q$) in the second term of the curly bracket, eq. (D.5).
Thus, a successive integration over  $d^D r$ leads to
\ba
W_{3-gluon}& =& -i\alpha \sum_{i=1}^3 \int \frac{d^D p \, d^Dq\,
e^{2ip_+T} }{(2\pi)^{2D}} \tilde{\cal F}_i (p, q)
\frac{(e^{2ip_-L} -1)(e^{2iq_-L} -1)}{p_-^2 q_- p^2 q^2 (p+q)^2 }
\left[1+e^{-2iL(p_-+q_-)} \right]  \nonumber\\ \noalign{\vskip 0.2cm}
&\equiv& \sum_{i=1}^3 W^{(i)}
\ea
where the amplitudes $\tilde{\cal F}_1 (p,q)$ are now defined by
\ba
\tilde{\cal F}_1 (p,q)&=&\frac{2}{[p_+]}
\left\{ \frac{1}{[q_+]} \{q_- (2p^2 + pq)
- p_- (q^2 +2pq)\}\right.\nonumber \\ \noalign{\vskip 0.2cm} &&\left.
 + \frac{1}{[p_+ + q_+]} \{q_- (pq - p^2) + p_- (pq -
q^2)\} \right\} \, , \\ \noalign{\vskip 0.2cm}
\tilde{\cal F}_2 (p,q)&=&\frac{2p_-}{[p_+]}(p_- + 2 q_-)\,, \\
\noalign{\vskip 0.2cm}
\tilde{\cal F}_3 (p,q)&=&\frac{2}{[p_+ + q_+]}\{p_-^2 - q_-^2\} -
\frac{2}{[q_+]}
\{q_- (q_- + 2q_-)\}\,.
\ea
To cast $\tilde{\cal F}_1$ in the
form (D.7), we repeatedly used the so called ``splitting formula''
\be
\frac{1}{[q_+][p_++q_+] } =\frac{1}{[p_+]} \left(
\frac{1}{[q_+] } -\frac{1}{[p_++q_+]}\right) \,,
\ee
which is an identity, in the sense of the theory of distributions,
when the spurious singularities are defined by ML  prescription.
Eq. (D.6) is particularly convenient for the
computation of the  three contributions $W^{(i)}$ in momentum
representation. In fact, for each of the terms $W^{(i)}$, it is necessary
 to compute only the term with the ``1'' factor in the last square
bracket of eq. (D.6), as the second one, proportional to $\exp [-2iL
(p_-+q_-)]$, can be obtained from the first by the replacement $L\to -L$.
Following these suggestions and using (for instance) Schwinger
parameterization for Feynman denominators and ML-distributions,
one can get the following results for $W^{(1)}$ and
$W^{(2)}$
\ba
W^{(1)} &=&-i\alpha\frac{(4LT)^{4-D} e^{-i\pi D/2} \cos (\pi D/2)}{
(2\pi)^D} \frac{\Gamma(D-4)}{(D-4)^2} \nonumber \\ \noalign{\vskip 0.2cm}
&&\times \left\{3 \frac{\Gamma (D/2-1) \Gamma(D/2 -2)}{\Gamma(D-2)}
+4\frac{\Gamma^2 (3-D/2)}{(D/2-1)\Gamma (5-D)} \right.\nonumber\\
\noalign{\vskip 0.2cm}
&&\left.
+4\Gamma(2-D/2)\Gamma(D/2-1)\frac{}{} \right\}\\ \noalign{\vskip 0.2cm}
W^{(2)}&=&i\alpha\frac{(4LT)^{4-D} e^{-i\pi D/2} \cos (\pi D/2)}{
(2\pi)^D} \frac{\Gamma(D-4)}{(D-4)^2} \nonumber \\ \noalign{\vskip 0.2cm}
&&\times \left\{\frac{\Gamma(D/2 -1) \Gamma(D/2 -2)}{\Gamma (D-2)}
-2\Gamma (D/2 -2) \Gamma (3-D/2) \right\}\,.
\ea
It should be stressed that, contrary to what happens in the coordinate
representation, the expressions $W^{(1)}$ and $W^{(2)}$
 have been obtained  keeping the
dimension $D$ in the natural strip $3<\Re D<4$, and therefore they do not
entail any analytical continuation.

The computation of $W^{(3)}$ is more
delicate and require additional care.
Following the same instructions used for $W^{(1)}$ and $W^{(2)}$, one can get
the following integral form for $W^{(3)}$
\ba
W^{(3)} &=&
i\alpha\frac{(4LT)^{4-D} e^{-i\pi D/2} \cos (\pi D/2)}{
(2\pi)^D} \frac{\Gamma(D-4)}{(D-4) }
\int_0^1 d\zeta \zeta^{D/2-2} \, \frac{2-\zeta}{1-\zeta}
\, \varphi (\zeta)
\nonumber \\ \noalign{\vskip 0.2cm}
&&\times\int_0^\infty \frac{d\eta}{1+\eta} [\eta +1-\zeta]^{D-5}
\bigl[ (\eta +1-\zeta)^{2-D/2} - (\eta\zeta)^{2-D/2} \bigr]\,,
\ea
where $\varphi (\zeta) = [\zeta^{4-D}  -(1-\zeta)^{4-D} +1 -2\zeta]$.
The integral in (D.13) presents nasty singularities in $\zeta =1$ if
$D$ lies in the natural strip.
To overcome this problem yet remaining in the $3<\Re D<4$ region,
we can for instance perform first an
integration by parts in $d\eta$ and write
\ba
\int_0^\infty \frac{d\eta}{1+\eta} (\eta +1-\zeta)^{D/2 -3}&=&
\frac{(1-\zeta)^{D/2-2}}{2-D/2} \nonumber
\\ \noalign{\vskip 0.2cm} &&- \frac{_2F_1 (2-D/2 , 3-D/2 ;
4-D/2;\zeta)}{(2-D/2)(3-D/2)}\,, \\ \noalign{\vskip 0.2cm}
\int_0^\infty \frac{d\eta}{1+\eta} (\eta +1-\zeta)^{D -5}
\eta^{2-D/2}&=&\frac{\Gamma^2(3-D/2)}{\Gamma (6-D)} (1-\zeta)^{D/2
-2}\nonumber\\ \noalign{\vskip 0.2cm} &&
\times {_2F_1 (1, 3-D/2; 6-D;\zeta)} \,.
\ea
Notice that the two hypergeometric series in the r.h.s. are analytic
functions in $|\zeta|\le 1$ if $D$ lies in its natural strip. Therefore,
the remaining integral in $d\zeta$ can be evaluated integrating term by
term in the series expansion of the hypergeometrics.
The final result is
\ba
W^{(3)}&=&i\alpha\frac{(4LT)^{4-D} e^{-i\pi D/2} \cos (\pi D/2)}{
(2\pi)^D} \frac{\Gamma(D-4)}{(D-4)^2}
 \left\{ \frac{1}{D/2-2} \left[\frac{\Gamma^2(3-D/2)}{\Gamma
(5-D)}\right.\right.
\nonumber \\ \noalign{\vskip 0.2cm}
&&\left. - 3 \Gamma(3-D/2) \Gamma(D/2 -1) + 2 \frac{\Gamma^2 (D/2
-1)}{\Gamma(D-2)}\right]
+ \frac{2}{\Gamma(2-D/2)}\sum_{n=0}^{\infty} \frac{\Gamma
(n+2-D/2)}{(n+3-D/2) n!}
 \nonumber\\ \noalign{\vskip 0.2cm}
&& \times\left[ \psi(n+D/2) - \psi(n+3-D/2) +
\frac{2}{(n+D/2-1)(n+D/2)}
\right.\nonumber\\ \noalign{\vskip 0.2cm}
&&\left. + \frac{1}{n+3-D/2} -
\frac{\Gamma(n+D/2-1)\Gamma(5-D)}{\Gamma(4+n-D/2)} \right]
+(4-D)\Gamma(3-D/2)\nonumber\\ \noalign{\vskip 0.2cm}
&&\times\sum_{n=0}^\infty \frac{\Gamma(n+3-D/2)}{\Gamma(n+6-D)} \left[
\Gamma(D/2-2)\left(\frac{\Gamma(n+5-D)}{\Gamma (n+3-D/2)} -
\frac{\Gamma (n+2)}{\Gamma (n+D/2)} \right) \right. \nonumber \\
\noalign{\vskip 0.2cm}
&&+ \left. \left. 2\frac{\Gamma(n+1)\Gamma (D/2)}{\Gamma(n+1+D/2)}
+\frac{\Gamma(n+5-D)\Gamma(D/2-1)}{\Gamma (n+4-D/2)}
-\frac{\Gamma(n+1)\Gamma(3-D/2)}{\Gamma (n+4-D/2)}\right]\right\}
\ea
where $\psi(z) =d\,  \log \Gamma (z)/dz$.
Eq. (D.16) concludes the calculation of $W^{(3)}$ and therefore of the whole
three-gluon graph. Notice that all the series in eq. (D.16) are
convergent for $3<\Re D<4$.

Although not straightforward, it can be checked
that collecting all the expressions for $W^{(i)}$ and performing the
Laurent expansion around $D=4$, the residues at the $D=4$ poles as well
as the
finite term exactly coincide with the corresponding quantities evaluated
in the coordinate representation, eq. (4.12).

\newpage
\bb{99}
\bi{Mig}  A.A.Migdal, Phys. Rep. 102 (1983) 316.
\bi{Pol}  A.M.Polyakov, Nucl. Phys. B164 (1980) 171;
\\        I.Ya.Aref'eva, Phys. Lett. B93 (1980) 347;
\\        V.S.Dotsenko and S.N.Vergeles, Nucl. Phys. B169 (1980) 527;
\\        R.A.Brandt, F.Neri and M.-A.Sato, Phys. Rev. D24 (1981) 879.
\bi{WL}   I.A.Korchemskaya and G.P.Korchemsky, Phys. Lett. 287B (1992) 169.
\bi{IR}   S.V.Ivanov, G.P.Korchemsky and A.V.Radyushkin,
          Sov. J. Nucl. Phys. 44 (1986) 145;
          G.P.Korchemsky and A.V.Radyushkin,
          Phys. Lett. 171B (1986) 459; Nucl. Phys. 283B (1987) 342.
\bi{KM}   G.P.Korchemsky and G.Marchesini,
          Parma Univ. preprint UPRF--92--354 [hep-ph/9210281].
\bi{KR}   G.P.Korchemsky and A.V.Radyushkin, Phys. Lett. 279B (1992) 359.
\bi{Geo}  For a review see: H.Georgi, ``Heavy Quark Effective
          Field Theory,'' preprint HUTP--91--A039 (1991).
\bi{RG}   G.P.Korchemsky, Mod. Phys. Lett. A4 (1989) 1257.
\bi{pres} 
          S.Mandelstam, Nucl. Phys. B213 (1983) 149;
          G.Leibbrandt, Phys. Rev., D29 (1984) 1699.
\bi{cons} 
          A.Bassetto, M.Dalbosco, I.Lazzizzera and R.Soldati,
          Phys. Rev., D31 (1985) 2012.
\bi{AT}   
          A.Andra\v{s}i and J.C.Taylor, Nucl. Phys. B375 (1992) 341.
\bi{gen}  
          A.Bassetto, M.Dalbosco and R.Soldati,
          Phys. Rev., D36 (1987) 3138;
          see also A.Bassetto, G.Nardelli and R.Soldati,
          ``Yang--Mills Theories in Algebraic Non-Covariant Gauges'',
          World Scientific, Singapore, 1991.
\bi{NET}  J.G.M.Gatheral, Phys. Lett. 133B (1984) 90;
\\        J.Frenkel and J.C.Taylor, Nucl. Phys. B246 (1984) 231.
\bi{prop} A.Bassetto, Phys. Rev. D46 (1992) 3676.
\bi{ghost} A.Bassetto, {\it in\/} Proceedings of the 8th Warsaw Symposium
          on Elementary Particle Physics, Kazimierz (Poland) 1985.
\bi{trans}M.Dalbosco, Phys. Lett. 180B (1986) 121.
\bi{Z}    A.Bassetto, M.Dalbosco and R.Soldati,
          Phys. Rev., D33 (1986) 617.
\eb
\newpage
\begin{center}
\def\S#1{{\scriptstyle {#1}}}
\unitlength=0.75mm
\linethickness{0.4pt}
\noindent
\begin{picture}(48.00,38.00)
\put(5.00,5.00){\vector(0,1){15.01}}
\put(5.00,5.00){\line(0,1){30.00}}
\put(5.00,35.00){\vector(1,0){19.97}}
\put(5.00,35.00){\line(1,0){40.00}}
\put(45.00,35.00){\vector(0,-1){15.01}}
\put(45.00,35.00){\line(0,-1){30.00}}
\put(45.00,5.00){\vector(-1,0){20.03}}
\put(45.00,5.00){\line(-1,0){40.00}}
\put(2.00,38.00){\makebox(0,0)[cc]{$\S{0}$}}
\put(48.00,38.00){\makebox(0,0)[cc]{$\S{n^\star}$}}
\put(48.00,2.00){\makebox(0,0)[cc]{$\S{n+n^\star}$}}
\put(2.00,2.00){\makebox(0,0)[cc]{$\S{n}$}}
\end{picture}
\par\bigskip\par
Fig.~1
\vspace*{15mm}
\par\bigskip\par
\noindent
\begin{picture}(110.00,55.00)
\put(67.00,22.00){\vector(0,1){15.01}}
\put(67.00,22.00){\line(0,1){30.00}}
\put(67.00,52.00){\vector(1,0){19.97}}
\put(67.00,52.00){\line(1,0){40.00}}
\put(107.00,52.00){\vector(0,-1){15.01}}
\put(107.00,52.00){\line(0,-1){30.00}}
\put(107.00,22.00){\vector(-1,0){20.03}}
\put(107.00,22.00){\line(-1,0){40.00}}
\put(64.00,55.00){\makebox(0,0)[cc]{$\S{0}$}}
\put(110.00,55.00){\makebox(0,0)[cc]{$\S{n^\star}$}}
\put(110.00,19.00){\makebox(0,0)[cc]{$\S{n+n^\star}$}}
\put(64.00,19.00){\makebox(0,0)[cc]{$\S{n}$}}
\bezier{24}(87.00,22.00)(87.00,37.00)(87.00,52.00)
\put(87.00,55.00){\makebox(0,0)[cc]{$\S{x_1(s)}$}}
\put(87.00,19.00){\makebox(0,0)[cc]{$\S{x_2(t)}$}}
\put(10.00,22.00){\vector(0,1){15.01}}
\put(10.00,52.00){\vector(1,0){19.97}}
\put(50.00,52.00){\vector(0,-1){15.01}}
\put(50.00,22.00){\vector(-1,0){20.03}}
\put(10.00,22.00){\line(0,1){30.00}}
\put(10.00,52.00){\line(1,0){40.00}}
\put(50.00,52.00){\line(0,-1){30.00}}
\put(50.00,22.00){\line(-1,0){40.00}}
\put(7.00,55.00){\makebox(0,0)[cc]{$\S{0}$}}
\put(53.00,55.00){\makebox(0,0)[cc]{$\S{n^\star}$}}
\put(53.00,19.00){\makebox(0,0)[cc]{$\S{n+n^\star}$}}
\put(7.00,19.00){\makebox(0,0)[cc]{$\S{n}$}}
\bezier{16}(20.00,52.00)(20.00,42.00)(30.00,42.00)
\bezier{16}(30.00,42.00)(40.00,42.00)(40.00,52.00)
\put(20.00,55.00){\makebox(0,0)[cc]{$\S{x_1(s_2)}$}}
\put(40.00,55.00){\makebox(0,0)[cc]{$\S{x_1(s_1)}$}}
\put(30.00,12.00){\makebox(0,0)[cc]{(a)}}
\put(87.00,12.00){\makebox(0,0)[cc]{(b)}}
\end{picture}
\par
Fig.~2
\vspace*{15mm}
\par\bigskip\par
\noindent
\begin{picture}(165.00,50.00)
\bezier{14}(12.00,40.00)(12.00,49.00)(21.00,49.00)
\bezier{14}(21.00,49.00)(30.00,49.00)(30.00,40.00)
\bezier{14}(21.00,40.00)(21.00,31.00)(30.00,31.00)
\bezier{14}(30.00,31.00)(39.00,31.00)(39.00,40.00)
\put(5.00,40.00){\vector(1,0){20.04}}
\put(5.00,40.00){\line(1,0){40.00}}
\bezier{16}(65.00,40.00)(65.00,50.00)(75.00,50.00)
\bezier{16}(75.00,50.00)(85.00,50.00)(85.00,40.00)
\bezier{8}(75.00,50.00)(75.00,45.00)(75.00,40.00)
\put(55.00,40.00){\vector(1,0){20.04}}
\put(55.00,40.00){\line(1,0){40.00}}
\bezier{16}(115.00,40.00)(115.00,50.00)(125.00,50.00)
\bezier{16}(125.00,50.00)(135.00,50.00)(135.00,40.00)
\put(105.00,40.00){\vector(1,0){20.01}}
\put(105.00,40.00){\line(1,0){40.00}}
\bezier{14}(125.00,22.00)(125.00,31.00)(125.00,40.00)
\put(5.00,45.00){\makebox(0,0)[cc]{$\S{0}$}}
\put(45.00,45.00){\makebox(0,0)[cc]{$\S{n^\star}$}}
\put(25.00,17.00){\makebox(0,0)[cc]{(a)}}
\put(55.00,45.00){\makebox(0,0)[cc]{$\S{0}$}}
\put(95.00,45.00){\makebox(0,0)[cc]{$\S{n^\star}$}}
\put(75.00,17.00){\makebox(0,0)[cc]{(b)}}
\put(105.00,45.00){\makebox(0,0)[cc]{$\S{0}$}}
\put(145.00,45.00){\makebox(0,0)[cc]{$\S{n^\star}$}}
\put(125.00,17.00){\makebox(0,0)[cc]{(c)}}
\end{picture}
\par
Fig.~3
\vspace*{15mm}
\par\bigskip\par
\noindent
\begin{picture}(162.00,55.00)
\put(5.00,22.00){\vector(0,1){15.01}}
\put(5.00,52.00){\vector(1,0){19.97}}
\put(45.00,52.00){\vector(0,-1){15.01}}
\put(45.00,22.00){\vector(-1,0){20.03}}
\put(5.00,22.00){\line(0,1){30.00}}
\put(5.00,52.00){\line(1,0){40.00}}
\put(45.00,52.00){\line(0,-1){30.00}}
\put(45.00,22.00){\line(-1,0){40.00}}
\put(2.00,55.00){\makebox(0,0)[cc]{$\S{0}$}}
\put(48.00,55.00){\makebox(0,0)[cc]{$\S{n^\star}$}}
\put(48.00,19.00){\makebox(0,0)[cc]{$\S{n+n^\star}$}}
\put(2.00,19.00){\makebox(0,0)[cc]{$\S{n}$}}
\put(62.00,22.00){\vector(0,1){15.01}}
\put(62.00,52.00){\vector(1,0){19.97}}
\put(102.00,52.00){\vector(0,-1){15.01}}
\put(102.00,22.00){\vector(-1,0){20.03}}
\put(62.00,22.00){\line(0,1){30.00}}
\put(62.00,52.00){\line(1,0){40.00}}
\put(102.00,52.00){\line(0,-1){30.00}}
\put(102.00,22.00){\line(-1,0){40.00}}
\put(59.00,55.00){\makebox(0,0)[cc]{$\S{0}$}}
\put(105.00,55.00){\makebox(0,0)[cc]{$\S{n^\star}$}}
\put(105.00,19.00){\makebox(0,0)[cc]{$\S{n+n^\star}$}}
\put(59.00,19.00){\makebox(0,0)[cc]{$\S{n}$}}
\put(119.00,22.00){\vector(0,1){15.01}}
\put(119.00,52.00){\vector(1,0){19.97}}
\put(159.00,52.00){\vector(0,-1){15.01}}
\put(159.00,22.00){\vector(-1,0){20.03}}
\put(119.00,22.00){\line(0,1){30.00}}
\put(119.00,52.00){\line(1,0){40.00}}
\put(159.00,52.00){\line(0,-1){30.00}}
\put(159.00,22.00){\line(-1,0){40.00}}
\put(116.00,55.00){\makebox(0,0)[cc]{$\S{0}$}}
\put(162.00,55.00){\makebox(0,0)[cc]{$\S{n^\star}$}}
\put(162.00,19.00){\makebox(0,0)[cc]{$\S{n+n^\star}$}}
\put(116.00,19.00){\makebox(0,0)[cc]{$\S{n}$}}
\bezier{30}(15.00,22.00)(25.00,37.00)(35.00,52.00)
\bezier{30}(15.00,52.00)(25.00,37.00)(35.00,22.00)
\bezier{8}(82.00,22.00)(82.00,27.00)(82.00,32.00)
\bezier{8}(82.00,52.00)(82.00,47.00)(82.00,42.00)
\bezier{8}(82.00,42.00)(77.00,42.00)(77.00,37.00)
\bezier{8}(77.00,37.00)(77.00,32.00)(82.00,32.00)
\bezier{8}(82.00,42.00)(87.00,42.00)(87.00,37.00)
\bezier{8}(82.00,32.00)(87.00,32.00)(87.00,37.00)
\bezier{12}(139.00,37.00)(139.00,30.00)(139.00,22.00)
\bezier{14}(139.00,37.00)(134.00,44.00)(129.00,52.00)
\bezier{14}(139.00,37.00)(144.00,44.00)(149.00,52.00)
\put(15.00,55.00){\makebox(0,0)[cc]{$\S{x_1(s_2)}$}}
\put(35.00,55.00){\makebox(0,0)[cc]{$\S{x_1(s_1)}$}}
\put(35.00,19.00){\makebox(0,0)[cc]{$\S{x_2(t_2)}$}}
\put(15.00,19.00){\makebox(0,0)[cc]{$\S{x_2(t_1)}$}}
\put(82.00,55.00){\makebox(0,0)[cc]{$\S{x_1(s)}$}}
\put(82.00,19.00){\makebox(0,0)[cc]{$\S{x_2(t)}$}}
\put(128.00,55.00){\makebox(0,0)[cc]{$\S{x_1(s_3)}$}}
\put(149.00,55.00){\makebox(0,0)[cc]{$\S{x_1(s_2)}$}}
\put(139.00,19.00){\makebox(0,0)[cc]{$\S{x_2(s_1)}$}}
\put(25.00,11.00){\makebox(0,0)[cc]{(a)}}
\put(82.00,11.00){\makebox(0,0)[cc]{(b)}}
\put(139.00,11.00){\makebox(0,0)[cc]{(c)}}
\put(141.00,36.00){\makebox(0,0)[cc]{$\S{z}$}}
\end{picture}
\par
Fig.~4
\par\bigskip\par
\end{center}
\newpage

{\Large {\bf Figure captions:}}

\bigskip

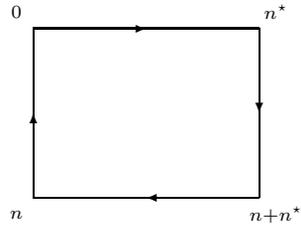
{\bf Fig.~1:} Integration path for Wilson loop. Parallel segments belong
on the light-cone.

\bigskip

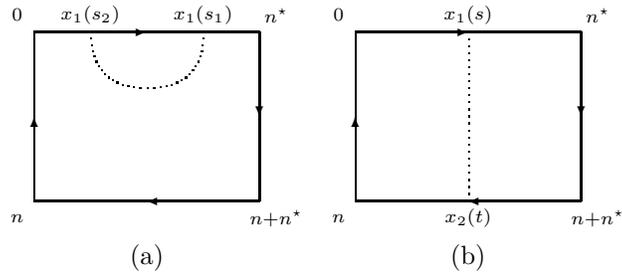
{\bf Fig.~2:} One-loop contributions to the Wilson loop expectation
value. The graph (a) vanishes due to causal properties of ML
prescription.

\bigskip

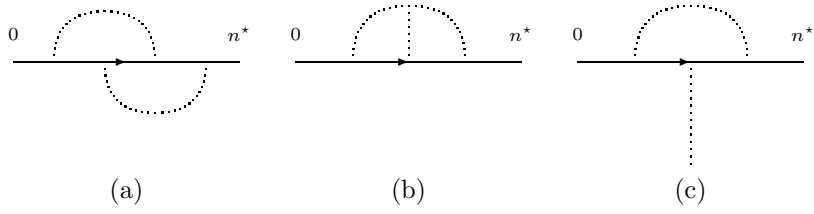
{\bf Fig.~3:} Diagrams containing the following subgraphs give
vanishing contributions to Wilson loop expectation value.

\bigskip

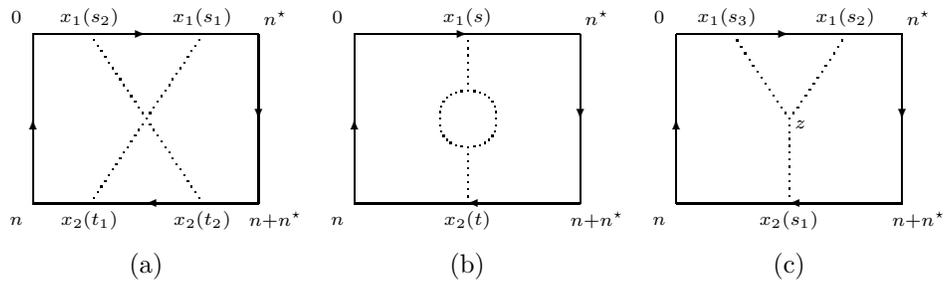
{\bf Fig.4:} Relevant diagrams contributing to the Wilson loop
to the second order: ``crossed'' diagram (a), ``self-energy'' diagram
(b) and ``three-gluon'' diagram (c).

\end{document}